\documentclass[
reprint,
amsmath,amssymb,
prl
]{revtex4-2}
\usepackage[utf8]{inputenc}
\usepackage{tikz}
\usepackage[export]{adjustbox}
\usepackage[normalem]{ulem}
\usepackage{physics}
\usepackage{graphicx,graphics}
\usepackage{dcolumn}
\usepackage{bm}
\usepackage{color}
\usepackage{lipsum}
\usepackage[percent]{overpic}
\usepackage{array}
\usepackage{nicefrac}
\usepackage{multirow}
\usepackage{latexsym,verbatim}
\usepackage{amsmath,amssymb,amsfonts}
\usepackage{xcolor}
\usepackage[breaklinks=true,colorlinks,citecolor=blue,linkcolor=blue,urlcolor=blue]{hyperref}
\makeatletter
\renewcommand{\thefigure}{\arabic{figure}}       
\renewcommand{\fnum@figure}{Figure~\thefigure}   
\makeatother

\begin{document}

\newcommand{\Ts}{T_{\rm S}}
\newcommand{\Tx}{T_{\rm 2DEG}}
\newcommand{\Tb}{T_{\rm b}}
\newcommand{\Qgis}{Q_{\rm GIS}}
\newcommand{\R}{\mathcal{R}}
\newcommand{\red}{\textcolor{red}}
\newcommand{\blue}{\textcolor{blue}}

\title{Strong nonlinear thermoelectricity generation and close-to-Carnot efficient heat engines in Superconductor-Insulator-2D electron gas junctions.}
\author{Leonardo Lucchesi}
\email{leonardo.lucchesi@unipi.it}
\affiliation{Dipartimento di Fisica, Università di Pisa, Largo Bruno Pontecorvo 3, 56127 Pisa, PI, Italy}
\affiliation{INFN Sezione di Pisa, Largo Bruno Pontecorvo 3, 56127 Pisa, PI, Italy}
\author{Federico Paolucci}
\affiliation{Dipartimento di Fisica, Università di Pisa, Largo Bruno Pontecorvo 3, 56127 Pisa, PI, Italy}
\affiliation{INFN Sezione di Pisa, Largo Bruno Pontecorvo 3, 56127 Pisa, PI, Italy}

\begin{abstract}
We propose and theoretically analyse a novel Superconductor-Insulator-2D electron gas tunnel junction (SI2DEG) that strongly
and efficiently generates thermoelectricity via a nonlinear mechanism. By varying the position of the electrochemical potential of the 2DEG, the SI2DEG junction shows different thermoelectric generation regimes with performance exceeding state-of-the-art tunnel systems. Indeed, the generated Seebeck potential can reach $6.75\Delta_0/e$ with a nonlinear Seebeck coefficient as high as $\mathcal{S}=5\,\Delta_0/e{\rm K}$. When operated as a quantum heat engine, the system efficiency gets very close to the Carnot limit with a maximum value $\eta=0.92\eta_C$ with a non-negligible power output. The SI2DEG junction also shows peculiar features such as bidirectional cooling controlled by the potential bias and bistability of the thermoelectric generation. These features, together with its strong temperature response, make the SI2DEG system an interesting general platform for quantum science and technology  applications, such as quantum thermodynamics, ultrasensitive cosmology and dark‑matter detection, and qubit refrigeration.
\end{abstract}
\maketitle
One of the main current limits for scaling cryogenic solid-state quantum technologies  \cite{Acin2018} is heat management. Indeed, the design of a scalable quantum computer must move from needing a coaxial cable per qubit towards the use of cryogenic electronics to reduce complexity and heat influx \cite{Savin2006,Tian2025,Xue2021,Bohuslavskyi2024,Sadhu2025,Kawabata2026}. Furthermore, scaling an array of superconducting detectors brings an analogous problem, as they usually require separate wires for biasing and readout \cite{Heikkilae2018,Kuzmin2019,DeLucia2024}, with the need for readout multiplexing \cite{Battistelli2008,DeLucia2024}. Generating thermoelectricity in a cryogenic environment offers an unexpected path for solving these problems, as it can passively generate signals or power only by using a byproduct of quantum devices: heat. As a consequence, thermoelectricity in superconducting structures
turned from a curiosity into a promising framework for quantum devices \cite{Heikkilae2018,Marchegiani2020,Paolucci2023}, useful for both quantum computation \cite{Acin2018,Kjaergaard2020} and quantum detection \cite{Giazotto2006,Degen2017} technologies.\\
Thermoelectric quantum heat engines can generate non-cyclic, steady power via particle exchange \cite{Humphrey2005}. Their efficiency has been discussed, mostly agreeing that Carnot efficiency can only be reached at zero power via energy filtering of the exchanged particles via a $\delta$-shaped transmission function \cite{Mahan1996, Humphrey2002, Jordan2013, Yamamoto2015}. The problem of maximum efficiency at finite power has also been addressed, setting the Curzon-Ahlborn limit in the linear regime \cite{VandenBroeck2005}, and the Whitney limit in the nonlinear regime \cite{Whitney2014,Whitney2015,Mishra2024}, both implying energy selection of transmitted particles \cite{Yamamoto2015,Whitney2015}. Further discussion on overcoming these limits for specific systems is ongoing \cite{Yamamoto2015,Benenti2011,Lee2017,Popp2021,Ryu2022}.\\
For scalable cryogenic thermoelectricity, superconducting hybrid structures appear as the most promising framework  \cite{Ozaeta2014,Marchegiani2020,Bianco2024}. The breaking of e-h symmetry required for thermoelectricity can happen spontaneously, as in superconductor-insulator-superconductor (SIS) junctions \cite{Marchegiani2020}, or explicitly, as in ferromagnetic-insulator-superconductor (FIS) junctions \cite{Ozaeta2014}, fluxon-based \cite{Singh2024}, and GIS \cite{Bianco2024} junctions. The generated thermoelectricity can be linear \cite{Ozaeta2014,Bianco2024}, i.e., proportional to the temperature gradient across the structure, or nonlinear \cite{Lucchesi2025,Marchegiani2020,Bianco2024}. \\
\begin{figure}[b]
\centering
\includegraphics[width=\columnwidth]{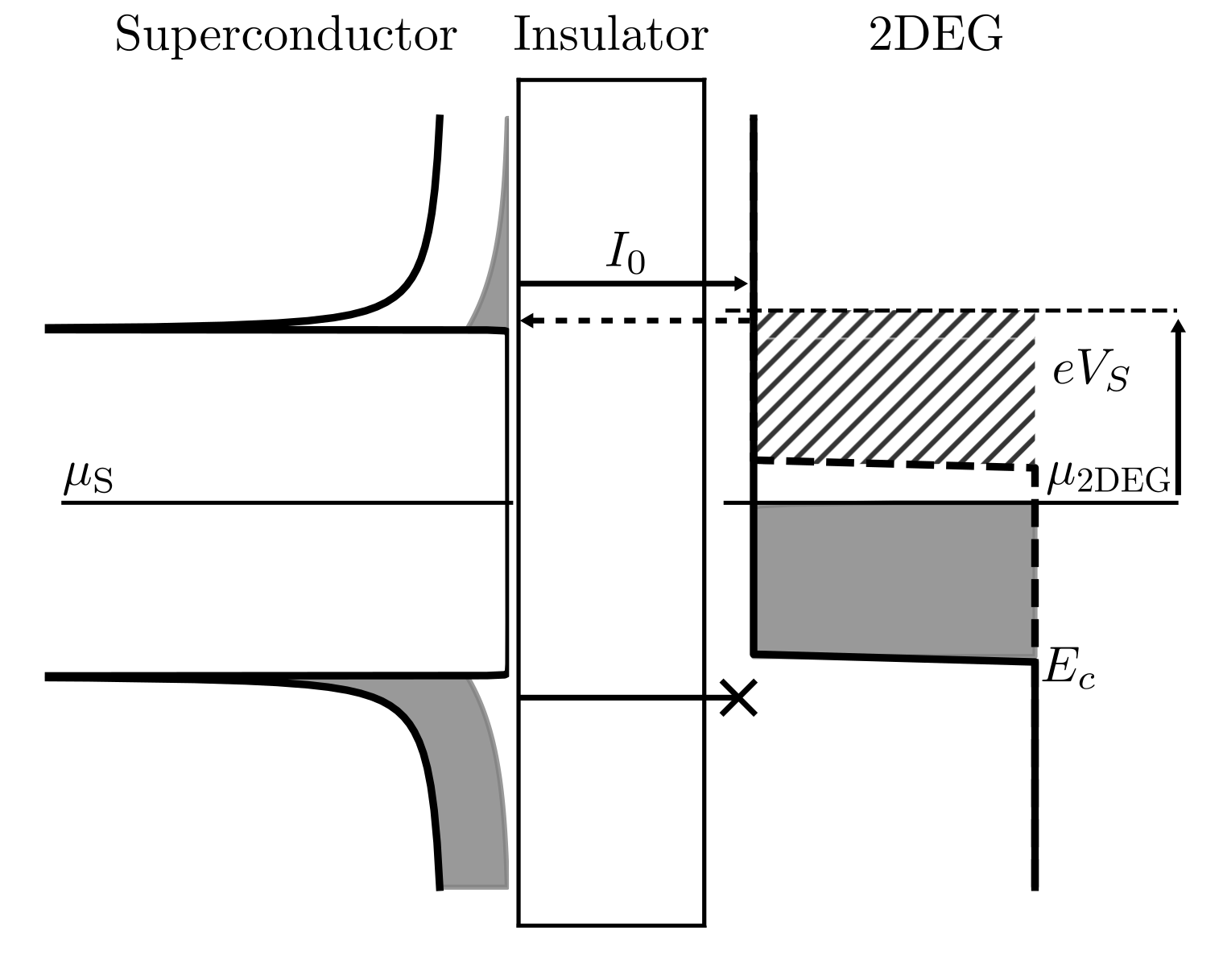}
\caption{SI2DEG junction scheme and band alignment for a moderately degenerate 2DEG ($E_c-\mu_{\rm 2DEG}=-0.8\Delta_0$). If $\Ts>\Tx$, a quasielectron current $I_0$ forms, while quasiholes are blocked by the 2DEG gap. In open-circuit, $V$ builds up until $\mu_{\rm 2DEG}$ reaches $\sim \Delta(T_{\rm S})$, where a current of electrons at the peak energy starts flowing backwards, canceling $I_0$. More information in the main text.}
\label{fig1}
\end{figure}
In this work, we introduce the Superconductor-Insulator-2D electron gas junction (SI2DEG) as a new system for cryogenic thermoelectricity generation that can reach record efficiency due to the energy filtering induced by the peak in the superconductor density of states (DOS) \cite{Mahan1996}. 
A SI2DEG junction addresses the main problems of current designs, as it does not need supercurrent suppression to generate a voltage gradient as a SIS junction \cite{Marchegiani2020,Germanese2022,Germanese2023}. After explaining the thermoelectricity generation mechanism, we evaluate the different thermoelectric regimes obtained by variation of band alignment. Finally, we discuss the main thermoelectric figures of merit, the peculiar bidirectional cooling and the potential applications of the Si2DEG system..\\
\textbf{Working principle and regimes \textendash}
The SI2DEG junction conducts current via quasiparticle tunneling across the insulator. 
The tunneling current is then described by a Landauer-like formula \cite{Tinkham}
\begin{multline}\label{eq:iv}
    I(V,\Ts,\Tx)=\frac{1}{eR_T}\int\,dE\; \rho_{\rm S}(E,\Ts)\cdot \\
    \cdot \rho_{\rm 2DEG}(E-eV)[f(E-eV,\Tx)-f(E,\Ts)],
\end{multline}
where $e$ is the electron charge, $V=(\mu_{\rm S}-\mu_{\rm 2DEG})/e$ is the applied potential across the junction, corresponding to the difference of superconductor and 2DEG chemical potentials $\mu_{\rm S},\mu_{\rm 2DEG}$ (we set $\mu_{\rm S}=0$), $\Ts$ and $\Tx$ are respectively the temperatures of the superconductor and the 2DEG and $f(E,T)$ is the Fermi function. $\rho_{\rm S}(E)=|{\rm Re}\, (E+i\Gamma)/\sqrt{(E+i\Gamma)^2-\Delta(\Ts)^2}|$ is the normalized DOS of the superconductor, with the Dynes parameter $\Gamma$ \cite{Dynes1984}.  $\rho_{\rm 2DEG}(E)=\theta(E-E_c)$ is the normalized DOS of the 2DEG, where $E_c$ is the position of the conduction band edge of the 2DEG with respect to $\mu_{\rm 2DEG}=0$. The tunneling resistance $R_T$ contains the normalization factors of both DOSs, the junction area and $\mathcal{T}$ \cite{Vischi2020}.\\
Fig.\ref{fig1} shows the band alignment of the SI2DEG junction for $E_c=-0.8\Delta_0$. When $\Ts>\Tx$, an excess of quasielectrons at energies $E\sim\Delta(\Ts)$ tunnels across the junction, while the 2DEG gap blocks tunneling of quasiholes at $E\sim-\Delta(\Ts)$. In a closed-circuit setup, this imbalance creates a Peltier current $I_0=I(V=0)$. In an open-circuit setup, this current charges the junction capacitance until the current stops flowing $I(V_S)=0$, defining the Seebeck potential $V_S$. This happens when the 2DEG chemical potential $\mu_{\rm 2DEG}$ is shifted over $\Delta(\Ts)$, allowing the tunneling from the more strongly occupied states close to $\mu_{\rm 2DEG}$. 
We now study thermoelectric generation in the junction by analyzing the $IV$ curves for different values of $E_c$ and $\Ts$, with $\Tx=7\times10^{-3}T_c$, obtaining different behaviors. We only analyze the effect of $E_c$ and $\Ts$, since $R_T$ only scales $I_0$. We set $\Gamma=10^{-4}\Delta_0$ for the rest of the article as it only introduce quantitative changes in the figures of merit, as shown in the SM \cite{SM}. We do not show the $\Tx>\Ts$ case as it does not show qualitative differences. For the open-circuit configuration, we consider a simplified electrical circuit where the junction acts as a current generator $I(V,\Ts,\Tx)$ with a capacitance $C$. We analyze the stability of every solution $I(V_0)=0$ by linearizing the circuit dynamics around $V_0$ as $\dot{v}=-(dI/dV|_{V_0})v/C$ \cite{Marchegiani2020}, where $v=V-V_0$. Every solution with $dI/dV|_{V_0}>0$ is stable.\\
\begin{figure}[tb]
\centering
\includegraphics[width=\columnwidth]{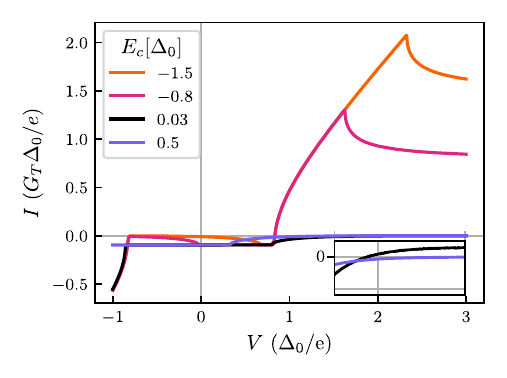}
\caption{IV curves for different values of $E_c$ at $T_{\rm S}=0.7T_c$. Thermoelectricity is generated in all regimes for $0\lesssim V\lesssim\Delta$, and for $V>\Delta$ if $E_c\geq 0$. The inset shows that for $V>\Delta$, $I$ changes sign for $E_c=0.03\Delta_0$, but not for $E_c=0.5\Delta_0$.}
\label{fig2}
\end{figure}
In general, all the $IV$ curves shown in Fig. \ref{fig2} include an area where $W=IV<0$, corresponding to the generation of thermoelectric power, induced by the temperature gradient $\Ts-\Tx=0.69 T_c$. For large negative $V$, the $IV$ curves show an Ohmic behavior, as expected, while for large positive $V$, they reach a constant value. Each result we show is also valid for a p-doped 2DEG, with appropriate sign corrections.\\
Now, we show the peculiarities of every different $E_c$ regime shown in Fig. \ref{fig2}.\\
\begin{figure*}[t]
\centering
\begin{tikzpicture}
  \node[anchor=south west, inner sep=0] (img) at (0,0)
    {\includegraphics[width=\textwidth]{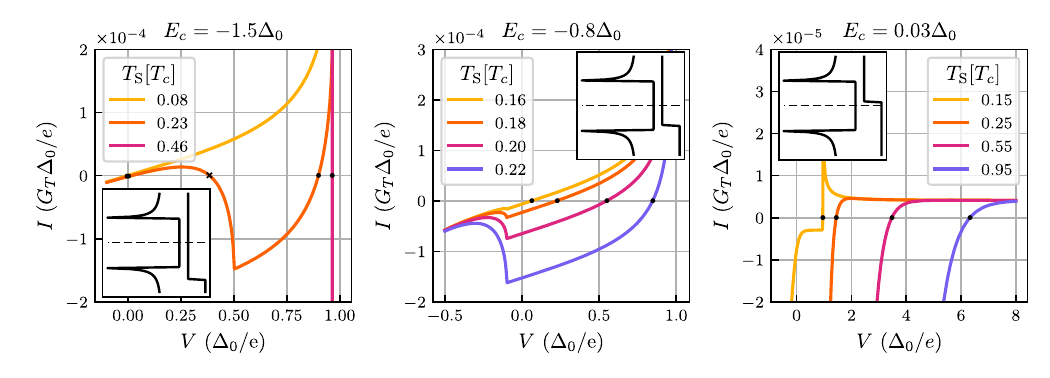}};
  \begin{scope}[x={(img.south east)}, y={(img.north west)}]
    \node[anchor=north west, color=black] at (0.0,0.94) {\textbf{a)}};
    \node[anchor=north west, color=black] at (0.36,0.94) {\textbf{b)}};
    \node[anchor=north west, color=black] at (0.68,0.94) {\textbf{c)}};
  \end{scope}
\end{tikzpicture}
\caption{Parts of IV curves for different values of $E_c$ and $\Ts$ showing different behaviors. \textbf{a)} $E_c=-1.5\Delta_0$. For different values of $\Ts$, the system transitions from an Ohmic-like behavior at $\Ts=0.08T_c$ to generating thermoelectricity at $\Ts=0.46T_c$. At $\Ts=0.23T_c$, the system shows an intermediate behavior that leads to a bistable state described in the main text.  \textbf{b)} $E_c=-0.8\Delta_0$. The system shows a strong dependence of $V_S$ on $\Ts$ for $0.15T_c\lesssim\Ts\lesssim0.22T_c$.  \textbf{c)} $E_c=0.03\Delta_0$. For $\Ts\gtrsim0.18T_c$, $V_S>\Delta$, up to $\sim 6\Delta_0$ for $\Ts=0.95T_c$.}
\label{fig3}
\end{figure*}
\textbf{$\mathbf{E_c<-\Delta_0}$ \textendash}
In this regime, the $IV$ curves show interesting behaviors only in the $0<V<\Delta_0/e$ interval, represented in Fig. \ref{fig3}a) . For $\Ts<0.22T_c$, we observe a dissipative behavior induced by the subgap states of the superconductor, while for $\Ts>0.25T_c$ the junction generates nonlinear thermoelectricity. $IV$ curves in the interval $0.22T_c<\Ts< 0.25T_c$ show two stable solutions $V\sim0,V\sim\Delta/e$ and an unstable one between them, a bistability that can be used as a thermal memory. 
The bistability range is quantitatively affected by the specific value of $E_c$ and $\Tx$, and by the value of $\Gamma$, as it increases the slope of the IV curve around $V=0$ and the width of the shoulder of $\rho_{\rm S}$ below $\Delta$. We show the effect of $\Gamma$ in the SM \cite{SM}.\\
\textbf{$\mathbf{-\Delta_0\leq E_c<0}$ \textendash}
This regime does not show bistability, generating a $I_0$ for every $\Ts>\Tx$. This regime shows a strong dependence of $V_S$ on $\Ts$ in the $0.15T_c\lesssim\Ts\lesssim 0.23T_c$ range, as shown in Fig. \ref{fig3}b). This is due to the transition from linear to nonlinear thermoelectricity generation, changing the main current contributor from subgap states to the peak in $\rho_{\rm S}$.\\
\textbf{$\mathbf{E_c\geq0}$ \textendash}
Fig. \ref{fig3}c) shows how IV curves change with $\Ts$ for the representative case $E_c=0.03\Delta_0$. For $\Ts<0.18T_c$, the system does not generate thermoelectricity, since the current rises above $0$ and stabilizes at an almost constant value. For $\Ts>0.18T_c$, the $IV$ curves cross zero at $V_S>\Delta(\Ts)/e$. The value of $V_S$ increases with temperature, as shown in Fig. \ref{fig3}c), reaching a maximum value of $V_S\sim6.75\Delta_0$ for $T\rightarrow T_c$. The reason for this behavior is that $\rho_{\rm 2DEG}$ only allows tunneling of the $E\gg0$ tails of the Fermi distributions. As increasing $V$ reduces the tunneled portion of $f(E,\Ts)$, $I$ changes sign when $f(E,\Ts)=f(E+eV,\Tx)$. As the tunneled portion of $f(E+eV,\Tx)$ does not depend on $V$, $V_S\propto \Ts$ for larger $\Ts$ where the larger $V_S$ implies $f(E,\Ts)\simeq \exp(-E/k\Ts)$. 
For $E_c\gtrsim0.4\Delta_0$, the $IV$ curve stops crossing $0$, as the cold distribution tail is cut off by $\rho_{\rm 2DEG}$ and we only see the exponential decay of the hotter tail. This is shown in the inset of Fig. \ref{fig2}.\\
\textbf{Thermoelectric figures of merit \textendash}
\begin{figure}[p]
\centering
\begin{tikzpicture}
  \node[anchor=south west, inner sep=0] (img) at (0,0)
    {\includegraphics[width=\columnwidth]{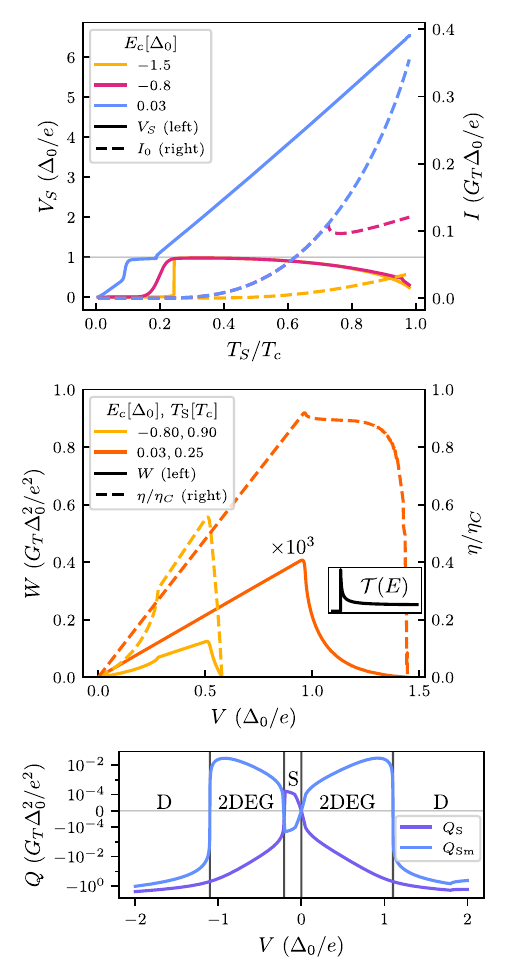}};
  \begin{scope}[x={(img.south east)}, y={(img.north west)}]
    \node[anchor=north west, color=black] at (0.0,1.0) {\textbf{a)}};
    \node[anchor=north west, color=black] at (0.0,0.63) {\textbf{b)}};
    \node[anchor=north west, color=black] at (0.0,0.24) {\textbf{c)}};
  \end{scope}
\end{tikzpicture}
\caption{
\textbf{a)} Dependence of Seebeck potential $V_S$ and Peltier current $I_0$ on $\Ts$ for different $E_c$. For $\!E_c\!<\!-\Delta_0$ and $-\Delta_0\!<\!E_c\!<\!0$, $V_S\sim\Delta_0/e$ even for relatively small $\Ts$. For $E_c\!>\!0$, $V_S\propto \Ts$ up to $\sim 6.75\Delta_0$, explained in main text. General behavior of $I_0$ and the peak at $\Ts\sim 0.75T_c$ are explained in the main text.
 \textbf{b)} Generated power $W$ and efficiency $\eta$ for different values of $E_c,\Ts$.
For $\Ts=0.25T_c$, $W$ is multiplied by $10^3$ for readability. $\eta_C$ is the Carnot efficiency. For $E_c=-0.8\Delta_0$, setting $\Ts=0.9T_c$ shows $W_{\rm max}=0.1\,G_T(\Delta_0/e)^2$. $E_c=0.03\Delta_0$ shows an efficiency of  $\eta=0.96\eta_C$ for $\Ts=0.25T_c$. Inset: transmission function. More information in the main text. For both \textbf{a)} and \textbf{b)}, $\Tx=7\times10^{-3}T_c$.
 \textbf{c)} Dependence of energy flow out of superconductor $Q_S$ and 2DEG $Q_{\rm 2DEG}$ on $V$ for $E_c=-0.8\Delta_0$ and $\Ts=\Tx=0.3T_c$. The system shows bidirectional cooling, with the presence of three regimes: superconductor cooling (S), 2DEG cooling (2DEG) and dissipative regime (D). 
}
\label{fig4}
\end{figure}
Fig. \ref{fig4}a) shows the dependence of $V_S$ and $I_0$ on $\Ts$ for the different $E_c$ regimes. For $\!E_c\!<\!\Delta_0$ $(-1.5\Delta_0)$ and $\!-\Delta_0\!<\!E_c\!<\!0$ $(-0.8\Delta_0)$, $V_S$ reaches $\sim\Delta_0/e$ for relatively small temperatures $\sim 0.25T_c$ and then decays as $\Delta(\Ts)$, with an uptick on the $-0.8\Delta_0$ curve at $\Ts=T^\ast$ where $2\Delta(T^\ast)=0.8\Delta_0$. For $E_c>0$ $(0.03\Delta_0)$, we see the $V_S\propto\Ts$ behavior, with $V_S$ reaching up to $6.75\Delta_0/e$. For this case, we numerically estimate a maximum nonlinear Seebeck coefficient $\mathcal{S}=V_S/(\Ts-\Tx)$ of $\mathcal{S}\sim 5\Delta_0/eK$. \\
The behavior of $I_0$ for $E_c=-1.5\Delta_0$ and $0.03\Delta_0$ are qualitatively similar, with $I_0$ increasing superlinearly with $\Ts$, with $E_c=-1.5\Delta_0$ showing smaller currents for the negative contribution of the superconductor quasiholes. $E_c=-0.8\Delta_0$ shows a hybrid behavior, as it behaves as $E_c=0.03\Delta_0$ until $\Ts^\ast\sim0.725T_c$, corresponding to $\Delta(\Ts^\ast)=|E_c|$. Here, the peak of $\rho_{\rm S}$ enters the conduction band of the 2DEG, allowing quasihole tunneling and thus reducing $I_0$.By further increasing $\Ts$, the system behaves as $E_c=-1.5\Delta_0$, as they differ only by a constant contribution.\\ 
Next, we analyze the generated work $W=-IV$ and the efficiency $\eta=W/Q_{\rm S}$ \cite{Marchegiani2020}, where $Q_{\rm S}$ is the total energy out from the hotter side (the superconductor) when the junction works as a heat engine and not as a heat pump. $Q_S$ is computed by multiplying the integrand in Eq.\ref{eq:iv} by $E/e$ (full expression in the SM \cite{SM}). We show $W$ and $\eta$ in Fig. \ref{fig4}c) as functions of $V$ for the $E_c=-0.8\Delta_0,\,\Ts=0.9 T_c$ and the $E_c=0.03\Delta_0,\,\Ts=0.25 T_c$ cases. We only plot the values of $V$ where the junction behaves as a heat engine, with positive $W>0$ and positive energy flow across the junction $Q_S>0$ (by convention, going from left to right). In the first case, we have a relatively large maximum power $W\sim0.1G_T(\Delta_0/e)^2$ at $V\sim\Delta(\Ts)/e$, similar to analogous junctions \cite{Marchegiani2020}, with an efficiency of $\eta\sim 0.55$. In the second case, we see a much smaller maximum $W\sim4\times 10^{-4}G_T(\Delta_0/e)^2$, related to the lower $\Ts$. However, the efficiency gets very close to the Carnot efficiency $\eta_{\rm max}=0.92\eta_C\sim0.89$, and it does it at the maximum of $W$. We also need to compare our $\eta$ to the Curzon-Ahlborn limit for efficiency at maximum $W$ in the linear regime $\eta_{\rm CA}=1-\sqrt{\Tx/\Ts}=0.826$ \cite{VandenBroeck2005}, and the equivalent Whitney limit $\eta_{\rm Wh}$ for the nonlinear regime \cite{Whitney2015}. We compute $\eta_{\rm Wh}=0.978\eta_C$ in the small $W_{\rm max}/W_{qb}$ approximation \cite{Whitney2015}, with $W_{qb}=0.0321\pi^2 Mk_B^2(\Ts-\Tx)^2/h$ and $M=h/2e^2R_T$ the effective number of modes, as $W_{\rm max}/W_{qb}\simeq0.0669$. Therefore, $\eta_{\rm max}>\eta_{\rm CA}$, higlighting the fundamental role of nonlinearity, and $\eta_{\rm max}=0.94\eta_{\rm Wh}$, implying that our system is very close to the maximum efficiency at maximum power. 
This very large efficiency can be ascribed to the energy filtering induced by the effective transmission probability $\mathcal{T}(E)= \rho_{\rm S}\rho_{\rm 2DEG}/(eR_T)$, shown in the inset of Fig. \ref{fig4}b). \\
The SI2DEG junction can be also exploited as a cooler. Fig. \ref{fig4}c) shows the voltage dependence of the heat flows coming out of the superconductor and the 2DEG, respectively $Q_S$ and $Q_{\rm 2DEG}$ for $\Ts=\Tx=0.3T_c$ and $E_c=-0.8\Delta_0$. The system shows both cooling of the superconductor $Q_S>0,\,Q_{\rm 2DEG}$ (S), and of the 2DEG $Q_S<0,\,Q_{\rm 2DEG}>0$ (2DEG). For larger $V$, the system starts to dissipate as expected, with $Q_S<0,\,Q_{\rm 2DEG}<0$ (D). Differently from conventional SIS junctions \cite{Manninen1999, Quaranta2011}, our SI2DEG system is capable to produce bidirectional cooling. At a given value of $E_c$, the direction of the cooling can be easily controlled by the voltage bias $V$, as shown in Fig. \ref{fig4}c).\\ 
\textbf{Conclusions~\textendash} In this Letter, we showed the physics of the strong thermoelectricity generation in a superconductor-insulator-2D electron gas (SI2DEG) junction and its potential for cryogenic applications. In separate cases, the junction generates an open-circuit potential up to $V_S\sim 6.75 \Delta_0/e$ ($1.35$ mV for Al) with maximum nonlinear Seebeck coefficient $\mathcal{S}\sim 1\,{\rm mV}/{\rm K}$ for Al, and a closed-circuit current up to $I_0 \sim0.12 G_T\Delta_0/e$, while the maximum values for a SIS junction are respectively $V_S\sim 0.16\,{\rm mV},\,\mathcal{S}\sim 0.3\,{\rm mV}/{\rm K}$, and $I_0=0$ \cite{Marchegiani2020}. As a heat engine, the junction is able to generate a power of $W\sim 0.1G _T(\Delta_0/e)^2$ ($\sim 4$ pW for Al with tunnel resistance $R_T=1\,{\rm k}\Omega$, the same as a SIS junction \cite{Marchegiani2020}). These thermoelectric figures of merit are similar or much better than analogous junctions \cite{Ozaeta2014,Marchegiani2020,Bianco2024,Singh2024}. The use of superconductors with higher $T_c$ improves $V_S,\, \mathcal{S}$ and $I_0$. The maximum efficiency is close to the Carnot efficiency $\eta=0.92\eta_C$, almost filling the Whitney nonlinear quantum limit $\eta=0.94\eta_{\rm Wh}$ and showing potential for quantum thermodynamics experiments. The junction also shows $V$-dependent bidirectional cooling, becoming the second tunnel system after the SIS junction to cool a superconductor, with no need of supercurrent suppression. Finally, the system also shows a bistable state in an interval of $\Ts$ for $E_c<-\Delta_0$ which could serve as a thermal memory or a thermal switch \cite{Marchegiani2020}, and two regimes where $V_S$ depends strongly on $\Ts$, the basis of a responsive thermometer \cite{Giazotto2006} or a sensitive radiation detector \cite{Golubev2001}.
\begin{acknowledgments}
The Italian Ministry of University and Research partially funded the work of L.L. and F.P. under the call PRIN2022 (Financed by the European Union – Next Generation EU) project EQUATE (Grant No. 2022Z7RHRS) and under the call FIS2 project QuLEAP (Grant No. FIS2023-00227). F.P. acknowledges the CSN V of INFN under the technology innovation grant STEEP for partial financial support.
\end{acknowledgments}

\bibliography{sism}

\clearpage 
\appendix
\onecolumngrid

\section*{Supplemental Material}
\subsection*{Full expression for heat transport}
As mentioned in the main text, the full expression for heat leaving the hotter superconductor side of the junction is
\begin{equation}\label{eq:heat}
    Q_{\rm S}(V,\Ts,\Tx)=\frac{1}{e^2R_T}\int\,dE\; E\rho_{\rm S}(E,\Ts) \rho_{\rm 2DEG}(E-eV)[f(E-eV,\Tx)-f(E,\Ts)],
\end{equation}
where $e$ is the electron charge, $V=(\mu_{\rm S}-\mu_{\rm 2DEG})/e$ is the applied potential across the junction, corresponding to the difference of superconductor and 2DEG chemical potentials $\mu_{\rm S},\mu_{\rm 2DEG}$ (we set $\mu_{\rm S}=0$), $\Ts$ and $\Tx$ are respectively the temperatures of the superconductor and the 2DEG and $f(E,T)$ is the Fermi function. $\rho_{\rm S}(E)=|{\rm Re}\, (E+i\Gamma)/\sqrt{(E+i\Gamma)^2-\Delta(\Ts)^2}|$ is the normalized DOS of the superconductor, with the Dynes parameter $\Gamma$. \\
When the junction works as a heat engine, this quantity represents the total heat $Q$ entering the heat engine. $Q$ is the quantity needed to compute the efficiency of the heat engine $\eta=W/Q$, where $W$ is the power produced.
\subsection*{Effect of the Dynes parameter $\Gamma$}
\begin{figure}[tbp]
\centering
\includegraphics[width=0.8\textwidth]{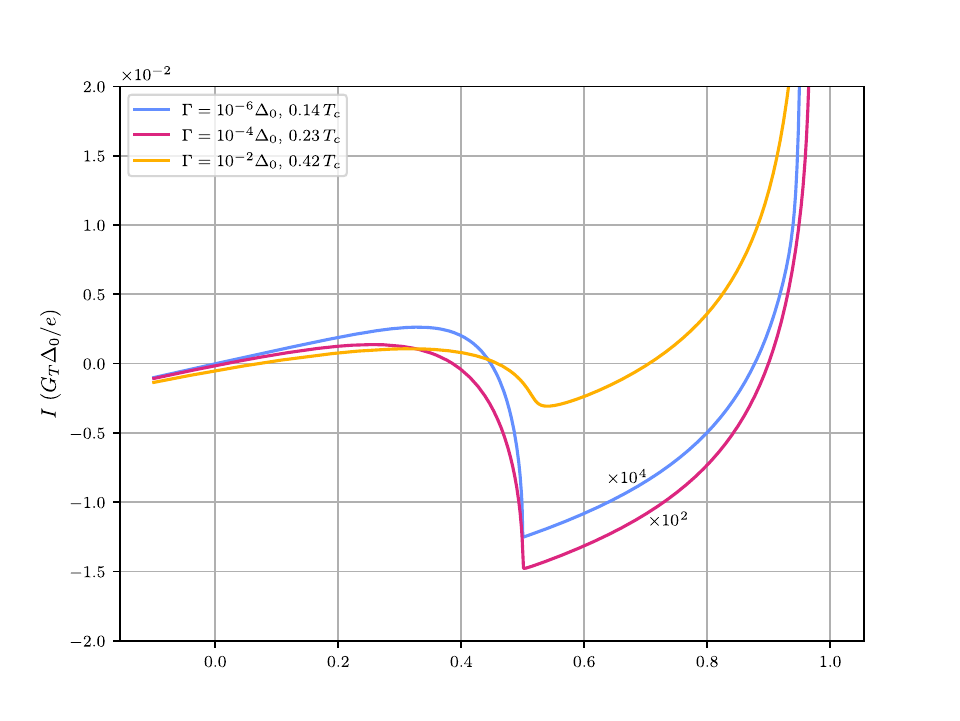}
\caption{$IV$ curves for different values of $\Gamma$ and $\Ts$, with $E_c=-1.5\Delta_0$ and $\Tx=7\times10^{-3}T_c$. The curves for $\Gamma=10^{-6}\Delta_0$ and $\Gamma=10^{-4}\Delta_0$ have been respectively multiplied by $10^4$ and $10^2$ for readability. $\Ts$ is chosen to have bistability for each value of $\Gamma$, with larger $\Gamma$ requiring larger $\Ts$ to reach bistability.}
\label{figs1}
\end{figure}

\begin{figure}[btp]
\centering
\begin{tikzpicture}
  \node[anchor=south west, inner sep=0] (img) at (0,0)
    {\includegraphics[width=0.8\textwidth]{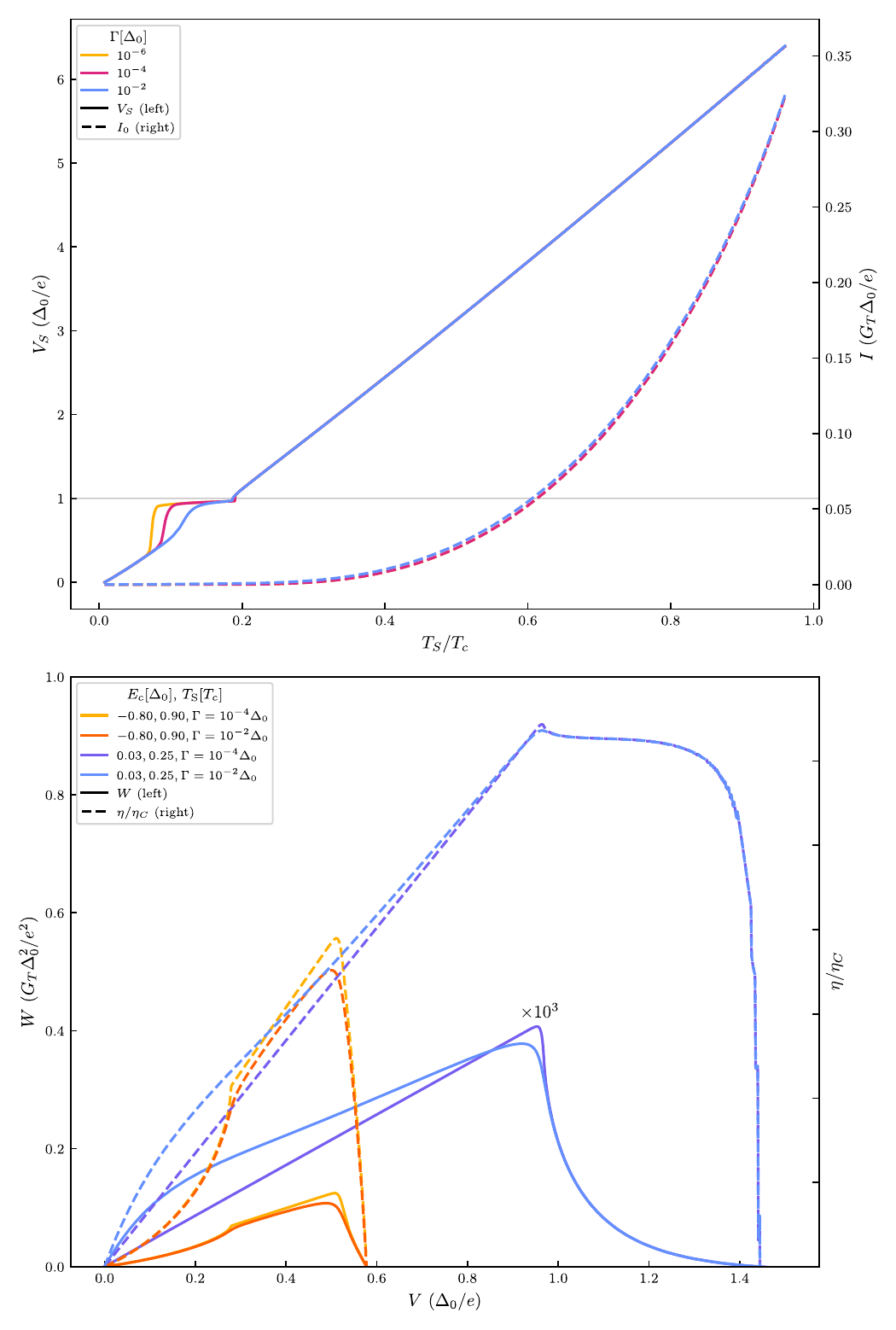}};
  \begin{scope}[x={(img.south east)}, y={(img.north west)}]
    \node[anchor=north west, color=black] at (0.0,1.0) {\textbf{a)}};
    \node[anchor=north west, color=black] at (0.0,0.5) {\textbf{b)}};
  \end{scope}
\end{tikzpicture}
\caption{
\textbf{a)} Dependence of Seebeck potential $V_S$ and Peltier current $I_0$ on $\Ts$ for different $E_c$ and $\Gamma$ for $E_c=0.03\Delta_0$. 
\textbf{b)} Generated power $W$ and efficiency $\eta$ for different values of $E_c,\Ts$ and $\Gamma$.
All figures of merit only show a quantitative dependence on $\Gamma$. 
}
\label{figs2}
\end{figure}

Here, we represent the dependence of the system behavior on the Dynes parameter $\Gamma$, representing the effective subgap density of states in the superconductor. As stated in the main text, the bistability pattern found for $E_c<\Delta_0$ shows a qualitative dependence on $\Gamma$. The bistability is due to the competition between the dissipative behavior induced by the subgap states and the thermoelectricity generated by the temperature gradient. Since $\Gamma$ represents the density of the subgap states, increasing $\Gamma$ shifts the balance towards dissipation, requiring a larger temperature gradient to have bistability. We show this effect in Fig. \ref{figs1}, where we represent $IV$ curves of the system for different values of $\Gamma$ and we set $\Ts$ to obtain bistability for each $\Gamma$. \\
In Fig. \ref{figs2}, we represent the main figures of merit for $\Gamma=10^{-4}\Delta_0$, representing a high quality superconductor and $\Gamma=10^{-2}\Delta_0$, representing a low quality superconductor. We can see that the main figures of merit only show a quantitative dependence on $\Gamma$. The Seebeck potential $V_S$ shows larger values for lower $\Gamma$ because a larger $\Gamma$ gives a larger shoulder of the superconductor DOS, stopping the thermoelectric charging of the junction for lower $V_S$. 

\end{document}